\documentclass[twocolumn,preprintnumbers,amsmath,amssymb,prb]{revtex4}
\usepackage{hyperref}
\usepackage{epsfig} 	
\usepackage{epsf}
\usepackage{dcolumn}
\usepackage{bm}

 \usepackage[latin1]{inputenc}

\def\dsize{\textstyle}

\def\dst #1{\displaystyle #1}
\def\MATH#1 {\begin{math}{\dst #1}\end{math} } 
\def\EQ#1 {\begin{equation}{\begin{array}{ll} \dst #1 \end{array}}\end{equation}}
\def\ft#1 {{\overvar{#1}{{}_{_{\dsize\hat{}}}}}}
\def\bs#1 {\boldsymbol #1}

\def\bold#1{\mbox{\bfseries #1}}
\def\mb#1{{\bold #1}}
\def\vb#1{\vec{\mb{#1}}}

\def\v#1{\vb{v}_#1}
\def\u#1{\vb{u}_#1}
\def\V{\vb{V}_{{\rm cm}}}
\def\q{\theta}

\begin{document}

\title{Geometric parametrization of binary elastic collisions}

\author{Amaro J. Rica da Silva}
\email{amaro@fisica.ist.utl.pt}
\author{Jos\'e P. S. Lemos}
\email{lemos@fisica.ist.utl.pt}
\affiliation{Centro Multidisciplinar de Astrofísica-CENTRA \\\&\\ Physics Dept., Instituto Superior Técnico, Universidade Técnica Lisboa\\ Av. Rovisco Pais, 1049-001 Lisbon, PORTUGAL}
\begin{abstract}
A geometric view of the possible outcomes of elastic collisions of two
massive bodies is developed that integrates laboratory, center of mass, and
relative body frames in a single diagram. From these diagrams all the
scattering properties of binary collisions can be obtained. The particular
case of gravitational scattering by a moving massive object corresponds to
the slingshot maneuver, and its maximum velocity is obtained.
\end{abstract}

\date{\today}
\maketitle

\section{Introduction}

We show how to geometrically parametrize the elastic collision 
between two bodies of arbitrary masses and velocities by using diagrams 
in the rest-frame of one of the bodies. For given 
masses and initial velocities the possible solutions in two dimensions can be parametrized by a single angle $\q$ for both attractive or repulsive interactions. 
Although an elastic collision is an highly idealized approximation of real interactions, there are
many applications where this approximation is appropriate.

The statement that two bodies collide means that for a very short
time $\delta t$ in comparison to the ratio of characteristic
length scales and speeds, the forces due to gravitational,
electromagnetic, or any other interaction between the two bodies dominate
any external forces in causing the momentum change $\Delta\vb{p}_{i}$ of each
of the bodies. This condition implies that the impulse received from the external forces
by one body during the collision is negligible in comparison to the impulse
contribution from the interaction forces due to the other body, which usually
justifies assuming that the total linear momentum or center of
mass momentum of the two bodies during the collision is a constant.
This assumption is an approximation because the total change of momentum of the 
system is equal to the total impulse
received. (The total impulse due to internal forces must add to zero because 
the forces on the two bodies must be instantaneously equal and opposite.) 
This approximation becomes
better as $\delta t\to 0$ or the external forces become weaker relative to the
internal forces and is exact when there are no external forces in which
case the center of mass momentum is a constant of the motion.

When the interactions are conservative, the total mechanical energy is conserved
during the collision, which leads to the assumption that the total kinetic energy
is conserved immediately before and after the collision, that is, when the internal
forces are (and again become) negligible in comparison to the external forces. 
This assumption is also an
approximation, which becomes asymptotically exact when there are no external forces.
When the internal forces are central, the total angular momentum is also conserved
in the collision under similar assumptions.

The usual treatment of the elastic collision of two bodies of masses $m_1$
and $m_2$ with initial velocities $\v0$ and $\u0$ invokes
conservation of linear momentum and kinetic energy in a one- or
two-dimensional setting.\cite{Armstr,Barger,Gold, Landau,Ramsey}
Three-dimensional collisions are seldom addressed, but see
Ref.~\onlinecite{Crawf}.

\begin{figure}[h]
   \centering
   \includegraphics[scale=.8]{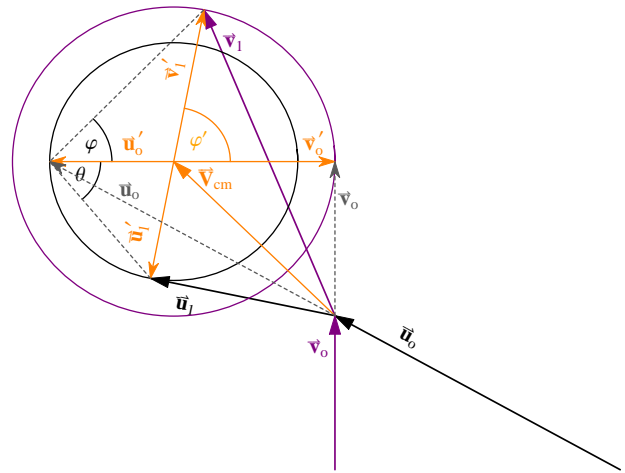}
   \caption{Laboratory and CM views of an elastic collision between different bodies of different masses ($m_2>m_1$) with initial velocities $\u0$ and $\v0$. When drawn from the same origin, $\V$ and the resulting velocities $\u1$ and $\v1$ always lie on a straight line, as also holds for $\V$ and $\u0$ and $\v0$. Furthermore, the possible outcomes of $\v1$ and $\u1$ lie on circumferences centered at $\V$ passing through $\v0$ and $\u0$. The angle $\varphi^{\prime}$ is the scattering angle in the CM frame and $\q$ and $\varphi$ represent the $u$- and $v$-scattering angles relative to the incoming velocity of the CM, viewed in the rest frame of the $u$-body.}
   \label{fig:LCM}
\end{figure}

 The view from the initial (rest) frame of one of the
bodies is usually worked out and related to the center of mass (CM) view of
the collision. The latter is particularly simple because the total linear
momentum
$\vb{P}_{{\rm cm}}$ is always zero in this reference frame, which means that
\begin{equation}
\v0^{\prime}=-\frac{m_2}{m_1} \u0^{\prime}\quad \mbox{and}\quad 
\v1^{\prime}=-\frac{m_2}{m_1} \u1^{\prime},
\label{eqn:cmm}
\end{equation}
where $\v{i}^{\prime}=\vb{v}_i-\vb{V}_{{\rm cm}}$ and
$\u{i}^{\prime}=\u{i}-\vb{V}_{{\rm cm}}$. Equation~(\ref{eqn:cmm}) means that
the incoming velocities appear as collinear opposing vectors in the CM frame
and so do the outgoing velocities. Conservation of kinetic energy is a
scalar equation which in the CM frame 
can be expressed as
\begin{equation}
|\v0^{\prime}|= |\v1^{\prime}|\quad \mbox{and}\quad
|\u0^{\prime}|=|\u1^{\prime}|.
\end{equation}

The resulting velocity directions remain undefined, so additional information
is necessary to completely determine the velocities, for example, the
scattering angle with respect to a reference direction. For a vector of
given magnitude but unknown direction the possible outcomes define the
points on a circumference centered on the origin. In the CM frame these vectors,
$\v1^{\prime}$ and
$\u1^{\prime}$, describe two circumferences whose diametrically opposed 
points represent the possible outcomes for the $v$-body and $u$-body velocities 
(see Fig.~\ref{fig:LCM}). To return to the laboratory frame it is necessary to 
add the constant vector
$\vb{V}_{{\rm cm}}$.

In the laboratory frame the following procedure can be used to geometrically
determine these outcomes. First notice that $\v1=\v0$ and
$\u1=\u0$ is also a possible solution, corresponding to a missed
collision. Extending $\v0$ and $\u0$ from the origin defines
points on two concentric circumferences whose center is pointed to by extending
$\vb{V}_{{\rm cm}}$ from the origin; these three points are in a
straight line. These two circumferences define all the possible velocities in the
laboratory frame, and once the direction of a resulting velocity is
determined, then so is the other by the collinearity of the three points (two
on each circumference plus the center). The collision trapezoid in
velocity space referred to in many textbooks is obtained in Fig.~\ref{fig:LCM} 
by joining all the arrowheads. 

These properties of the binary elastic collision are well known and will not be discussed further. Instead,
we will develop an alternative geometric interpretation of the conservation laws and relate the laboratory view, the CM view, and the view of the collision from
the initial reference frame of one of the bodies. This latter reference
frame is more practical because it is asymptotically coincident at
$t\to-\infty$ with the non-inertial body frame of the relative coordinates
and velocities for which the two-body problem for central forces is usually
solved; that is, the frame in which the relative motion of
the two masses will appear as that of a single reduced mass
$\mu =m_1 m_2/(m_1+m_2)$, at the relative position of one of the
masses, moving under forces pointing to a fixed total mass
$M=m_1+m_2$ at the frame's origin, where the other mass is at rest; 
this motion can be calculated for sufficiently well behaved forces. In
particular, for a gravitational collision the result must be a hyperbola
(unless the asymptotic provisos made in the preceding paragraphs do not apply,
for example if the two bodies do not start or end sufficiently far apart, 
and then elliptic and parabolic collisions should be considered).

\section{Elastic collision: Two-dimensional case with
one body at rest}\label{sec2}

In a collision with a mass at rest at the origin, the resulting trajectories
lie on a plane in which the total angular momentum $\vb{L}_0$
relative to the CM is orthogonal to both $\v1$ and $\u1$ (strong
action-reaction law).

\begin{figure*}[ht]
\includegraphics[scale=.67]{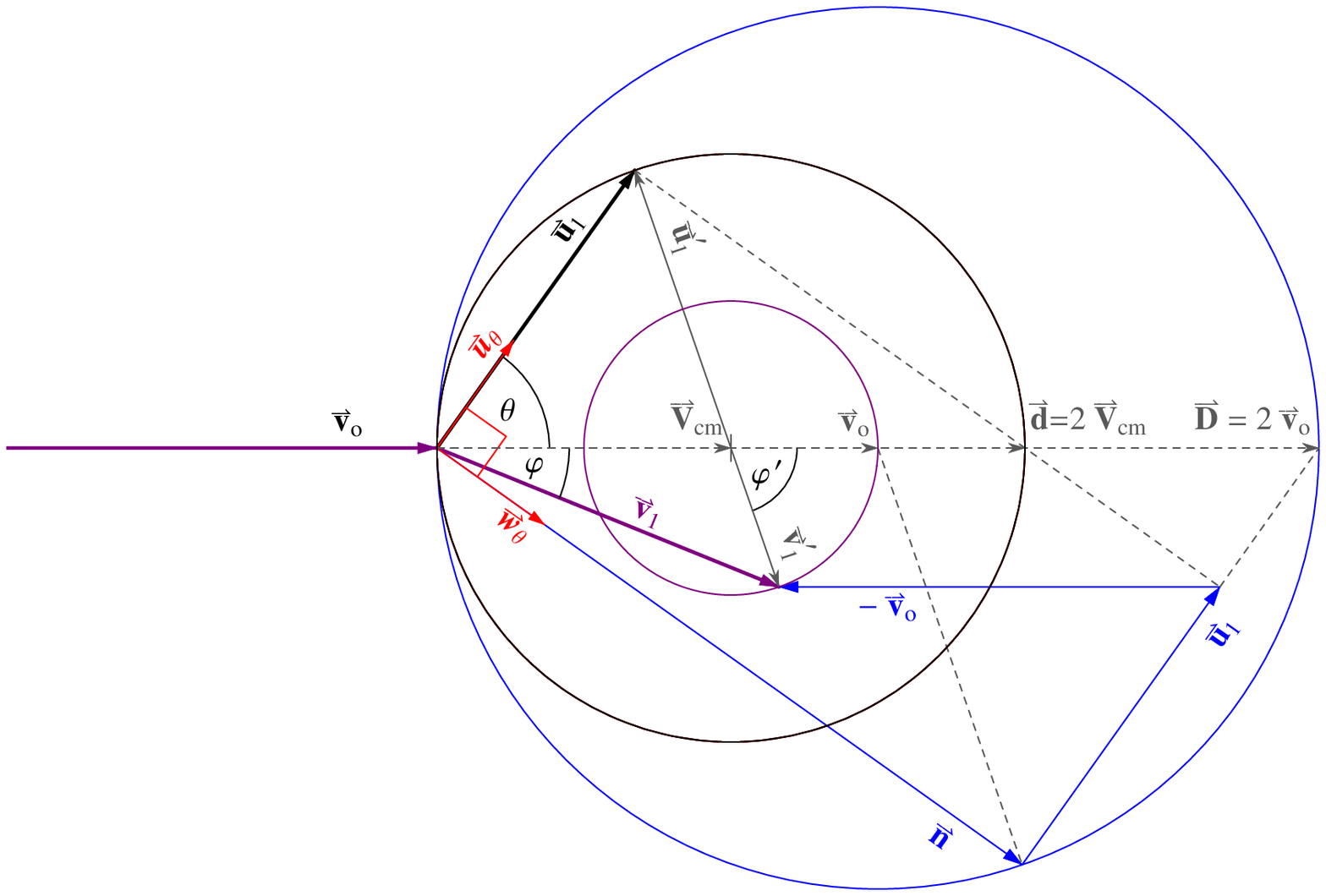}
\caption{Collision diagram for different masses $m_{1}>m_{2}$, $u$-body (mass $m_{2}$) at rest $\u0=0$. The circumference with diameter $\vb{d}=2\V$ is the locus of all possible outcomes $\u1$. As chords of this circumference, $\u{1}$ and $\vb{d}-\u{1}$ always subtend an angle $\pi/2$. A choice of $\q$ determines $\u1$ and $\vb{n}$, which is always orthogonal to $\u1$ and is restricted to a circumference of diameter $\vb{D}=2\v{0}$. The outcome velocity $\v1$ is geometrically determined by $\vb{n}+\u1-\v{0}$ and lies on a circumference centered on $\V$ with radius $|\v{0}-\V |$. In this diagram $\u1^{\,\prime}$ and $\v1^{\,\prime}$ represent the outcome velocities for the $u$- and $v$-bodies as seen from the CM frame, in which the incoming velocities would be $-\V$ and $\v{0}-\V$ respectively.}
\label{fig:different}
\end{figure*}

In the frame where the $u$-body is initially at rest, the relevant conservation equations for a collision with a \mbox{$v$-body} with velocity $\v{o}$ are
\begin{subequations}
\begin{align}
m_1\v0 & =m_1\v1 +m_2\u1 , \\
\frac{1}{2}m_1v_0^2 & =\frac{1}{2}m_1v_1^2+\frac{1}{2}m_2u_1^2,
\end{align}\label{eqn:PE}
\end{subequations}
which represent linear momentum conservation and kinetic energy conservation
before and after an elastic collision. An equivalent way of writing Eq.~(\ref{eqn:PE}) is
\begin{subequations}
\begin{align}
\frac{m_1}{m_2} (\v0-\v1)& =\u1 \label{eqn:LM} \\
\frac{m_1}{m_2}(\v0-\v1)\cdot (\v0+\v1)&=\u1\cdot \u1
.\label{eqn:KE}
\end{align}
\end{subequations}

We use Eq.~(\ref{eqn:LM}) to replace $m_1(\v0-\v1)/m_2$ by $\u1$ on
the left-hand side of Eq.~(\ref{eqn:KE}), and $\u1$ by
$m_1(\v0-\v1)/m_2$ on the right-hand side and obtain
\begin{equation}
\label{this}
\u1\cdot (\v0+\v1)=\u1\cdot
\Big(\frac{m_1}{m_2}(\v0-\v1)\Big).
\end{equation}

Equation~\eqref{this} can be solved to express the
unknown scalar product $\v1\cdot \u1$ in terms of
$\v0\cdot \u1$ as
\begin{equation}
\v1\cdot \u1=\frac{m_1-m_2}{m_1+m_2} \v0\cdot \u1.
\label{eqn:vou1}
\end{equation}
We take the scalar product of $\u1$ with both sides of 
Eq.~(\ref{eqn:LM}) and use Eq.~(\ref{eqn:vou1}) to eliminate $\v1\cdot \u1$ and obtain
\begin{equation}
\frac{2 m_1}{m_1+m_2}\v0\cdot \u1 =\u1\cdot \u1.
\label{eqn:uu1}
\end{equation}
Equation~\eqref{eqn:uu1} expresses the magnitude 
\begin{equation}
u_1=|\u1|=\sqrt{\u1\cdot \u1}\label{eqn:um}
\end{equation}
in terms of
the unknown angle
$\q$ that $\u1$ makes with $\v0$. We denote the outgoing direction of
the $u$-body by
\begin{equation}
\hat{\bf u}_\q=\frac{\u1}{u_1}.
\end{equation}
Then Eq.~(\ref{eqn:uu1}) is equivalent to
\begin{equation}
u_1=\frac{2 m_1}{m_1+m_2}\v0\cdot \hat{\bf u}_\q.
\end{equation}
Apart from the unknown value of $\q$, the resulting $u$-body velocity must
be
\begin{equation}
\u1=\frac{2 m_1}{m_1+m_2} v_0 \cos(\theta) \hat{\bf u}_\q.
\label{eqn:u1}
\end{equation}

Equation (\ref{eqn:LM}) can now be used to deduce an expression for $\v1$:
\begin{equation}
\v1=\v0-\frac{m_2}{m_1}\u1,
\label{eqn:v1}
\end{equation}
or in terms of $\q$,
\begin{equation}
\label{express}
\v1=\v0-\frac{2 m_2}{m_1+m_2} v_0 \cos(\q)\hat{\bf u}_\q.
\end{equation}
Equation~\eqref{express} is not particularly illuminating with regard
to its geometrical relation to $\u1$, so a more geometrical approach will be
adopted in Sec.~\ref{sec:geo} for the determination of $\v1$.

\subsection{Geometrical view}\label{sec:geo}

\begin{figure*}[ht]
\includegraphics[scale=.67]{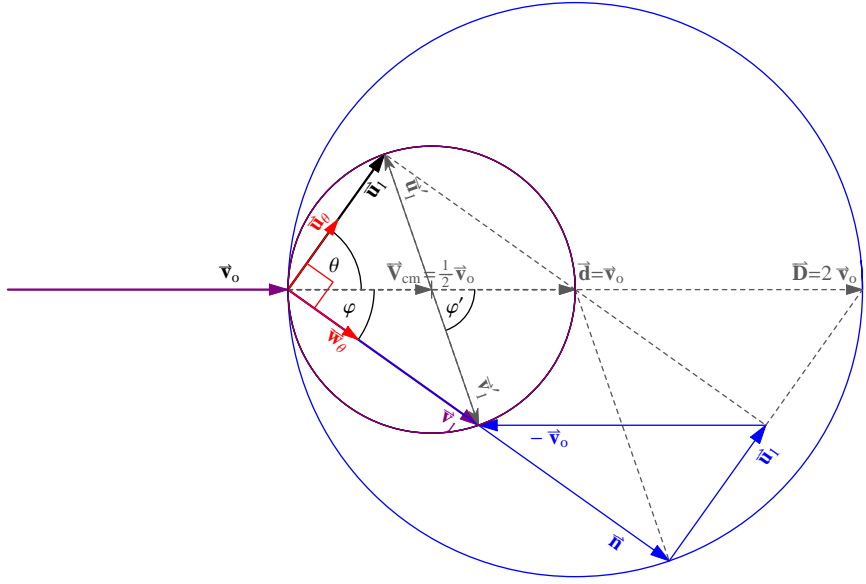}
\caption{Collision diagram for equal masses $m_{1}=m_{2}$, with the $u$-body (mass $m_{2}$) at rest ($\u0=0$). The $\u1$ and $\v1$ circumferences now coincide, and necessarily $\v1\,||\,\vb{n}\perp\u1$, that is $\q+\varphi=\frac{\pi}{2}$.}
\label{fig:equal}
\end{figure*}

Equation~(\ref{eqn:uu1}) can be expressed in terms
of the known vector $\vb{V}_{{\rm cm}}$. We set
\begin{equation}
\vb{d}=\frac{2 m_1}{m_1+m_2}\v0=2\vb{V}_{{\rm cm}}, 
\end{equation}
and collect terms on the left-hand side using the fact that $\vb{d}\cdot \u1=\u1\cdot \u1$, or
\begin{equation}
(\vb{d}-\u1)\cdot \u1=0,
\label{eqn:oc1}
\end{equation}
that is, $\vb{d}-\u1$ is always orthogonal to $\u1$. Equation~(\ref{eqn:oc1})
shows that any admissible solution $\u1$ will define a chord from the origin to 
a point on the circumference with fixed diameter defined by
$\vb{d}$ (see Fig.~\ref{fig:different}). 

The angle $\q \in [-\pi/2,\pi/2]$ formed by $\u1$ and
$\v0$ is the same as the angle between $\u1$ and $\vb{d}$. Thus, $\hat{\bf u}_\q$ is
the unit vector in the direction determined by $\q$, and for a
given choice of $\q$,
\begin{equation}
\u1=d \cos(\q)\,\hat{\bf u}_\q.
\end{equation}
Given the
geometric constraints on
$\u1$, we can determine
$\v1$ by substituting
$m_1(\v0-\v1)/m_2$ for $\u1$ on the left-hand side of
Eq.~(\ref{eqn:KE}). After rearranging terms we obtain
\begin{equation}
\label{state}
\u1\cdot (\v0+\v1-\u1)=0.
\end{equation}
Equation~\eqref{state} means that the vector $\vb{n}$ given by
\begin{equation}
\vb{n}=\v0+\v1-\u1,
\label{eqn:n}
\end{equation}
is always orthogonal to $\u1$,
\begin{equation}
\u1\cdot\vb{n}=0.
\label{eqn:un}
\end{equation}

We will show now that, just like $\u{1}$, the vector $\vb{n}$ is uniquely 
determined as soon as $\q$ is given. The expression for $\v1$ will then also 
be determined from Eq.~(\ref{eqn:n}) as
\begin{equation}
\v1=\u1-\v0+\vb{n}.
\label{eqn:vn}
\end{equation}

We use the orthogonality condition expressed by Eq.~(\ref{eqn:un}) to cancel the
right-hand side in the scalar product of Eq.~(\ref{eqn:LM}) with $\vb{n}$, 
 and replace Eq.~(\ref{eqn:vn}) on its left-hand side
to find $(m_1/m_2)\vb{n}\cdot (\v0-\v1)=0$, or
\begin{equation}
\vb{n}\cdot (2\v0-\vb{n})=0.
\label{eqn:oc2}
\end{equation}
That is, $\vb{n}$ and $\vb{D}-\vb{n}$ (where $\vb{D}=2\v0$) are always
orthogonal. Similar to Eq.~(\ref{eqn:oc1}), Eq.~\eqref{eqn:oc2} means that
$\vb{n}$ defines a chord from the origin to a point on the circumference with 
a fixed diameter defined by
$\vb{D}$. Because $\vb{n}$ is orthogonal to
$\u1$, this condition defines a unique chord that makes the angle
$\phi=(\pi/2)-\q$ with $\v0$. Its direction defines the unit vector
$\hat{\bf w}_\q$, orthogonal to $\hat{\bf u}_\q$, and therefore
\begin{equation}
\label{eq:this2}
\vb{n}=2 v_0 \cos\Big(\frac{\pi}{2}-\q
\Big)\hat{\bf w}_\q=2 v_0 \sin(\q)\hat{\bf w}_\q.
\end{equation}
We use Eqs.~\eqref{eq:this2} and (\ref{eqn:u1}) for $\u1$ and
decompose $\v0$ into
\begin{equation}
\v0=v_0 (\cos(\q)\hat{\bf u}_\q+\sin(\q)\hat{\bf w}_\q),
\label{eqn:vouw}
\end{equation}
so that Eq.~(\ref{eqn:vn}) becomes
\begin{equation}
\v1=v_0 \Big[\frac{m_1-m_2}{m_1+m_2} \cos(\q) \hat{\bf u}_\q+\sin(\q)
\hat{\bf w}_\q\Big].
\label{eqn:vuw1}
\end{equation}
The possible outcomes of $\v1$ also have a geometrical locus defined by a
circumference centered at
\begin{equation}
\vb{V}_{{\rm cm}}=\frac{1}{2}\vb{d}=\frac{m_1}{m_1+m_2}\vb{v}_0
\label{eqn:CM}
\end{equation}
away from the origin, with radius 
\begin{equation}
r_v=|\vb{v}_0-\vb{V}_{{\rm
cm}}|=\frac{m_2}{m_1+m_2}v_0.
\label{eqn:rv}
\end{equation}
This radius is to be expected because $\vb{v}_1=\vb{v}_0$, $\vb{u}_1=0$ 
is one possible result for the collision, meaning that the closest approach 
of the two bodies was too far compared to the range of the interaction forces. 
Confirmation that in general
the possible $\v1$ define such a circumference results from
verifying the orthogonality condition for two particular chords
\begin{equation}
(\v1-\v0)\cdot [\v1-\v0+2 (\v0-\vb{V}_{{\rm cm}})]=0,
\end{equation}
which holds when Eqs.~(\ref{eqn:CM}),
(\ref{eqn:vn}), and (\ref{eqn:LM}) are used, together with the identities
$\u1\cdot\vb{d} =\u1\cdot \u1$ and $\u1\cdot\vb{n}=0$:
\begin{equation}
(\v1-\v0)\cdot \Big(\u1-\frac{2 m_1 \v0}{m_1+m_2}+\vb{n}\Big)=
-\frac{m_2}{m_1} \u1\cdot (\u1-\vb{d}+\vb{n})=0.
\end{equation}

\subsection{Scattering angles in the $u$-body and center of mass frames}

\begin{figure*}[ht]
\includegraphics[scale=.67]{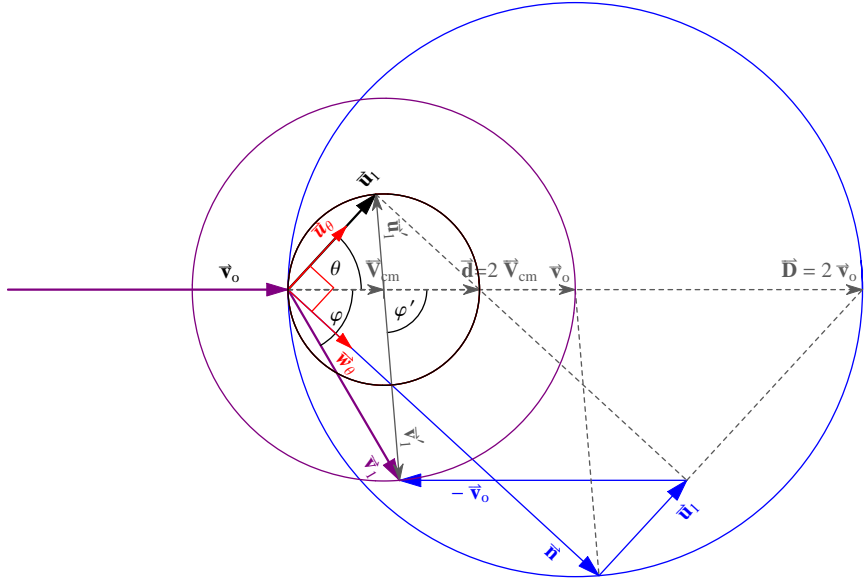}
\caption{Collision diagram for different masses $m_{1}<m_{2}$, $u$-body (mass $m_{2}$) at rest,
$\u0=0$. The $\v1$ circumference is greater than the $\u1$ circumference so back-scattering occurs when $|\varphi|>\frac{\pi}{2}$.}
\label{fig:mvlessmu}
\end{figure*}
The scattering angles for the collision can now be deduced from the 
parameters $m_1$, $m_2$, and $\q$. Relative to the invariant 
direction $\V$ of the CM, the scattering angle for $\u1$  is evidently 
$\q$ itself. As for the scattering angle $\varphi$ between $\v1$ and 
$\V \,||\,\v0$, it can be
calculated from Eqs.~(\ref{eqn:vouw}) and (\ref{eqn:vuw1})
\begin{equation}
\cos(\varphi)=\frac{\v1\cdot \v0}{v_1 v_0}=\frac{1-\frac{2
m_2}{m_1+m_2} \cos^2(\q)}{\sqrt{1-\frac{4 m_1
m_2}{(m_1+m_2)^2}\cos^2(\q)}}.
\label{eqn:qf}
\end{equation}
The total scattering angle between $\u1$ and $\v1$ is $\q+\varphi$.

If $m_1=m_2$, the identity (\ref{eqn:qf}) reduces to
$\cos(\varphi)=\sin(\q)$, which means that $\varphi=\phi=(\pi/2)-\q$
and
\begin{equation}
\u1\cdot\v1=u_1 v_1 \cos(\varphi+\q)=0.
\end{equation}
Thus, if
$m_1=m_2$, the resulting velocities $\u1$ and $\v1$ are orthogonal and
are both chords of the circumference with diameter $\vb{d}=\v0$ (see
Fig.~\ref{fig:equal}). This orthogonality is geometrically visible from
Eq.~(\ref{eqn:vou1}), which in this equal-mass case reduces to $\v1\cdot\u1=0$,
indicating that the solutions to the collision must remain orthogonal.

Note that if the angle $\varphi$ is specified instead of the angle $\q$,
the determination of $\u1$ is geometrically unique when $m_1<m_2$, but there
is an ambiguity when $m_1>m_2$ because there are two different
magnitudes for $\v1$ with the same $\varphi$, hence two different angles
$\q$. Direct inversion of Eq.~(\ref{eqn:qf}) yields
\begin{equation}
\cos^2 (\theta)=\frac{1}{2} \left[1+\frac{m_1}{m_2}\sin^2 (\varphi) \pm\cos
(\varphi) \sqrt{1-\frac{m_1^2}{m_2^2}\sin^2 (\varphi)}\right].
\label{eqn:tp}
\end{equation}

If $m_1<m_2$ the solution with the $-$ sign is the correct one. This
solution can be argued by looking at the limiting case of a missed
collision ($\varphi=0$) with the $m_2$ body at rest. In such a case $\theta$
is ill-defined (because the $m_2$ velocity remains zero), but we can
see that in neighboring collision cases, the limit of $\theta$ as
$\varphi\rightarrow\pm 0$ is $\pm \pi/2$, not zero (see
Fig.~\ref{fig:mvlessmu}). Actually $\theta=0$ for a head-on
collision, and because $m_1<m_2$ the outcome for $\v1$ corresponds to a
back-scattering with $\varphi=\pi$.

If $m_1>m_2$, real solutions exist only for
$\varphi\in[-\varphi_{L},\varphi_{L}]$ (indicating that
there is no back-scattering in these cases), where
\begin{equation}
\varphi_{L}=\pm\arcsin\Big(\frac{m_2}{m_1}\Big).
\end{equation}
The two solutions $\q_{\pm}(\varphi)$ obtained in Eq.~(\ref{eqn:tp}) now
apply. Geometrically this result could be obtained by referring to
Fig.~\ref{fig:different} and noting that the limiting values $\vb{v}_{1_L},
\varphi_{L}$ for $\v1$ and $\varphi$ are obtained when $\v1$ is
tangent to its locus circumference, that is, perpendicular to
$\v1^{\prime}=\v1-\V$. Because $\vb{v}_{1_L}\cdot\vb{v}_{1_L}^{\prime}=0$,
\begin{equation}
\vb{v}_{1_L}\cdot\V={|\vb{v}_{1_L}|}^2.
\end{equation}
Because ${|\vb{v}_{1_L}|}^2=V_{{\rm
cm}}^2-r_v^2=(m_1-m_2)/(m_1+m_2)v_0^2$, we have $\cos(\varphi_{L})=\sqrt{1-(m_2/m_1)^2}$, or
\begin{equation}
\sin({\varphi_{L}})=\pm
\frac{m_2}{m_1}.
\end{equation}

The scattering angle $\varphi $ for the $v$-body can also be related to the
scattering angle $\varphi^{\prime}$ as seen from the CM-frame. We use
$\v1=\v1^{\,\prime}+\V$ together with
Eqs.~(\ref{eqn:CM}) and (\ref{eqn:rv}) and obtain 
\begin{equation}
\tan(\varphi)=\frac{v_1^{\prime} \sin (\varphi')}{V_{{\rm cm}}+v_1^{\prime}
\cos (\varphi ')}=\frac{m_2 \sin (\varphi ')}{m_1+m_2 \cos (\varphi
')}.
\end{equation}
For equal masses
\begin{equation}
\tan(\varphi)=\tan\Big(\frac{\varphi '}{2}\Big).
\end{equation}

Likewise, the scattering angle $\q$ for the $u$-body can also be related to $\varphi^{\prime}$. If we use the fact
that $u_1^{\prime}= V_{{\rm cm}}$ and $\u1=\u1^{\prime}+\V$, we find\begin{equation}
\tan(\theta)=\frac{u_1^{\prime} \sin (\varphi')} {V_{{\rm cm}}-
u_1^{\prime} \cos (\varphi')}=\frac{\sin (\varphi ')}{1- \cos (\varphi ')},
\end{equation}
which simplifies to
\begin{equation}
\tan(\theta)=\cot (\frac{\varphi^{\prime}}{2}),
\end{equation}
which is independent of the mass ratio.

\begin{figure*}[ht]
\includegraphics[scale=.75]{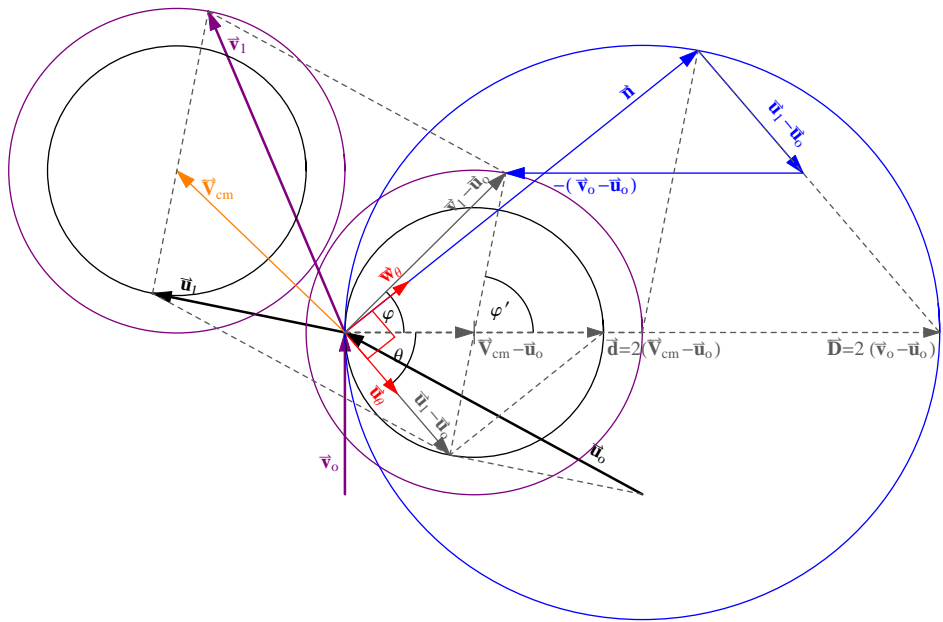}
\caption{Collision diagram for arbitrary masses $m_{1}< m_{2}$ with initial
velocities $\v0$, $\u0$ viewed in the Laboratory frame. The $u$-body rest frame diagram is used to obtain the final velocities $\u1$, $\v1$. The relation of this diagram to that represented in Fig.~\ref{fig:LCM} is also shown.}
\label{fig:general}
\end{figure*}

\section{Elastic Collision: General two-dimensional case}
\enlargethispage{4\baselineskip}

If both bodies are initially moving when viewed from a laboratory frame,
the same analysis can be carried out (see Fig.~\ref{fig:general}). All that
is necessary is to switch temporarily to an equivalent inertial frame
$S_{u_0}$ moving with the initial $u$-body velocity $\u0$. In
this frame the situation is exactly as before, that is, an elastic collision
with a body initially at rest. The same equations and conclusions are valid
except that everywhere we need to let $\v{i}\to\v{i}-\u0$,
$\u{i}\to\u{i}-\u0$, and $\vb{V}_{{\rm cm}}\to \vb{V}_{{\rm
cm}}-\u0$.

In this case the conservation equations are
\begin{subequations}
\begin{gather}
m_1\v0+m_2\u0=m_1\v1+m_2\u1;\\
\frac{1}{2}m_1v_0^2+\frac{1}{2}m_2u_0^2=\frac{1}{2}m_1v_1^2
+\frac{1}{2}m_2u_1^2,
\end{gather}
\end{subequations}
which can be rewritten as 
\begin{subequations}
\begin{gather}
\frac{m_1}{m_2}(\v0-\v1)=\u1-\u0
\label{eqn:lm2};\\
\frac{m_1}{m_2}(\v0-\v1)\cdot
(\v0+\v1)= (\u1-\u0)\cdot (\u1+\u0).
\label{eqn:ke2}
\end{gather}\label{eqn:LM1} 
\end{subequations}
Manipulation of Eq.~(\ref{eqn:LM1}) in a manner similar to Sec.~\ref{sec2}
will generate the equivalent relations in this new frame $S_{u_0}$.
Equation (\ref{eqn:uu1}) now becomes
\begin{equation}
|\u1-\u0|^2 = \frac{2 m_1}{m_1+m_2}(\v0-\u0)\cdot
(\u1-\u0).
\label{eqn:uuo1}
\end{equation}
We set as before
\begin{align}
\vb{d}&= 2(\vb{V}_{{\rm cm}}-\u0) = \frac{2
m_1}{m_1+m_2}(\v0-\u0) \\
\hat{\bf u}_\q&= \frac{1}{|\u1-\u0|}(\u1-\u0),
\end{align}
where
$\hat{\bf u}_\q$ is the direction of $\u1-\u0$ and $\q$ is the
angle between $\v0-\u0$ and $\u1-\u0$. Then
\begin{equation}
|\u1-\u0|= \vb{d}\cdot \hat{\bf u}_\q=\frac{2
m_1}{m_1+m_2}|\v0-\u0|\cos(\q).
\end{equation}
Note that Eq.~(\ref{eqn:uuo1}) is equivalent to 
\begin{equation}
(\u1-\u0)\cdot (\u1-\u0)= \vb{d}\cdot
(\u1-\u0),
\end{equation}
which means that
\begin{equation}
\label{eq:this3}
(\u1-\u0)\cdot (\u1-\u0-\vb{d})=0.
\end{equation}
As with Eq.~(\ref{eqn:oc1}), Eq.~\eqref{eq:this3} states that
$\u1-\u0$ always defines a chord from the origin to a circumference of diameter
$\vb{d}$. The final expression for $\u1$ is thus
\begin{equation}
\u1=\u0+\frac{2 m_1}{m_1+m_2}|\v0-\u0|\cos(\q)
\hat{\bf u}_\q.
\end{equation}
The equivalent of the $\vb{n}$ vector in Eq.~(\ref{eqn:n}) is
\begin{equation}
\vb{n}=(\v0-\u0)+\v1-\u1,
\label{eqn:n2}
\end{equation}
and its orthogonality to $\u1-\u0$ remains
\begin{equation}
(\u1-\u0)\cdot \vb{n}=0.
\label{eqn:oc3}
\end{equation}
Therefore Eq.~(\ref{eqn:oc2}) becomes
\begin{equation}
\vb{n}\cdot (2(\v0-\u0)-\vb{n})=0,
\label{eqn:oc4}
\end{equation}
and $\vb{n}$ also defines a chord from the origin to a circumference of diameter $\vb{D}=2(\v0-\u0)$. Thus geometrically once $\u1$ is defined then so is $\vb{n}$ and Eq.~(\ref{eqn:n2}) yields the final expression for $\v1$ as
\begin{equation}
\v1=\vb{n}+\u1- (\v0-\u0).
\label{eqn:v13}
\end{equation}

In this general case the only invariant direction in the collision is
that of the center of mass. It is left as an exercise for the reader to
derive a relation between $\q$, $\varphi$, and $\varphi^{\prime}$ in the
$u$-body and CM frames and the scattering deviations from the center of mass
direction of the resulting $\u1,\v1$ velocities in the lab frame.

\section{Conclusion}

We have shown how the relative velocities in a binary elastic collision
obey simple geometric relations even for arbitrary masses and
initial velocities. As can be seen from Fig.~\ref{fig:general}, the possible
final velocities
$\u1$ and $\v1$ for given initial conditions lie on two concentric circumferences
centered at a point in velocity space defined by $\vb{V}_{{\rm cm}}$. The points
defined by $\u1$, $\v1$, and $\vb{V}_{{\rm cm}}$ from the origin always
define a straight line. These circumferences have radii
\begin{subequations}
\begin{align}
r_u=|\u{1}-\vb{V}_{{\rm cm}}| & =\frac{m_1}{m_1+m_2}|\v0-\u0| \\
r_v=|\v{1}-\vb{V}_{{\rm cm}}| & =\frac{m_2}{m_1+m_2}|\v0-\u0|,
\end{align}
\end{subequations}
where $\v1-\u1$ was replaced by $\v0-\u0$ because, from
Eq.~(\ref{eqn:v13}) and the orthogonality condition (\ref{eqn:oc4}) (see
Fig.~\ref{fig:LCM}),
\begin{equation}
|\vb{v}_1-\vb{u}_1|^2 =
\vb{n}\cdot (\vb{n}-2 (\vb{v}_0-\vb{u}_0))+|\vb{v}_0
-\vb{u}_0|^2= |\vb{v}_0-\vb{u}_0|^2.
\end{equation}
The maximum radius for either of these circumferences is $r_i=|\v0-\u0|$,
which occurs when the respective $i$-body mass is much smaller
than the other, in which case the latter circumference will have a vanishing
diameter, meaning that the velocity of the massive body is similar to that
of the center of mass itself. If, for instance, $m_1\gg m_2$, then
$\vb{V}_{{\rm cm}}\approx\v0\approx\v1$. In the limit $m_2/m_1\to 0$,
$r_v\to 0$ and
\begin{equation}
r_u=\frac{m_1}{m_1+m_2} |\v0-\u0|\to |\v0-\u0|.
\end{equation}

All the scattering
properties for binary collisions can be obtained from collision diagrams such as those in Figs.~(\ref{fig:different})--(\ref{fig:general}). In particular, the
conditions for
$v$-body back-scattering or the existence of a maximum scattering angle
depends only on the condition
$r_v\gtrless r_u$. An interesting exercise would be to derive
the angular $\q$ range for which an increase in outgoing velocity is
obtained for a binary elastic collision with $m_1<m_2$, $\u0=0$.

One of us has written a Java application that renders these collision
diagrams interactively.\cite{Amaro} The program uses a Live Java library
developed by Martin Kraus.\cite{Kraus} Simulations
reveal scattering situations that are not intuitively obvious but can be
understood when the full two-body motion is explored. In particular, by
changing the asymptotic angle $\q$ and mass ratio
$m_2/m_1$, we can obtain the optimum incident condition for a
gravity-assisted fly-by (or gravitational slingshot).\cite{Broucke} It is then apparent
that the maximum velocity attainable by the smaller mass, $m_1< m_2$, in a
collision is 
\begin{equation}
v_{1,\, \max}=V_{{\rm cm}}+\frac{m_2}{m_1+m_2}|\v0-\u0|
\label{eqn:GSL}
\end{equation}
which occurs when both masses exit in the same direction as that of
$\vb{V}_{{\rm cm}}$. As a consequence of Eq.~(\ref{eqn:GSL}), when the
massive body is initially at rest (or when the collision is viewed from its
rest frame), the maximum velocity attained is
$v_{1,\,\max}=v_0$.

In future work we will show that these collision diagrams are useful for calculating the eccentricities and focal
distances for open Keplerian orbits for gravitational
scattering or repulsive Coulomb scattering. An explanation of the slingshot
maneuver and gravity-assist planetary fly-by can be easily obtained. In
this way the orbits can be viewed in the laboratory frame and a study can be
made of the optimal incidence angle for a planetary fly-by that delivers the
maximum velocity boost in a chosen direction.

\end{document}